\documentclass[sigconf]{acmart}

\AtBeginDocument{%
  }

\copyrightyear{2024}
\acmYear{2024}
\setcopyright{acmlicensed}\acmConference[VRST '24]{30th ACM Symposium on Virtual Reality Software and Technology}{October 9--11, 2024}{Trier, Germany}
\acmBooktitle{30th ACM Symposium on Virtual Reality Software and Technology (VRST '24), October 9--11, 2024, Trier, Germany}
\acmDOI{10.1145/3641825.3687708}
\acmISBN{979-8-4007-0535-9/24/10}

\acmConference[VRST 2024]{In 30th ACM Symposium on Virtual Reality
Software and Technology}{October 09--11,
  2024}{Trier, Germany}
\acmISBN{979-8-4007-0535-9/24/10}

\begin{document}

\title{Evaluating the Effects of Situated and Embedded Visualisation in Augmented Reality Guidance for Isolated Medical Assistance}

\author{Frederick G. VICKERY}
\affiliation{%
  \institution{Lab-STICC UMR 6285, ENIB}
  \city{Brest}
  \country{France}}
\email{frederickgvickery@gmail.com}

\author{Sébastien KUBICKI}
\affiliation{%
    \institution{Lab-STICC UMR 6285, ENIB}
    \city{Brest}
    \country{France}}
\email{sebastien.kubicki@enib.fr}

\author{Charlotte HOAREAU}
\affiliation{%
    \institution{Lab-STICC UMR 6285}
    \city{Brest}
    \country{France}}
\email{phd.charlotte.hoareau@gmail.com}

\author{Lucas BRAND}
\affiliation{%
    \institution{Lab-STICC UMR 6285, ENIB}
    \city{Brest}
    \country{France}}
\email{lucas.brand@enib.fr}

\author{Aurélien DUVAL}
\affiliation{%
    \institution{Lab-STICC UMR 6285, ENIB}
    \city{Brest}
    \country{France}}
\email{aurelien.duval@enib.fr}

\author{Seamus THIERRY}
\affiliation{%
    \institution{Groupe Hospitalier Bretagne Sud}
    \city{Lorient}
    \country{France}}
\email{seam.thi@gmail.com}

\author{Ronan QUERREC}
\affiliation{%
    \institution{Lab-STICC UMR 6285, ENIB}
    \city{Brest}
    \country{France}}
\email{ronan.querrec@enib.fr}

\renewcommand{\shortauthors}{Vickery et al.}

\begin{abstract}
  One huge advantage of Augmented Reality (AR) is its numerous possibilities of displaying information in the physical world, especially when applying Situated Analytics (SitA).
  AR devices and their respective interaction techniques allow for supplementary guidance to assist an operator carrying out complex procedures such as medical diagnosis and surgery, for instance.
  Their usage promotes user autonomy by presenting relevant information when the operator may not necessarily possess expert knowledge of every procedure and may also not have access to external help such as in a remote or isolated situation (\textit{e.g.}, International Space Station, middle of an ocean, desert).

  In this paper, we propose a comparison of two different forms of AR visualisation: An embedded visualisation and a situated projected visualisation, with the aim to assist operators with the most appropriate visualisation format when carrying out procedures (medical in our case). 
  To evaluate these forms of visualisation, we carried out an experiment involving 23 participants possessing latent/novice medical knowledge.
  These participant profiles were representative of operators who are medically trained yet do not apply their knowledge every day (\textit{e.g.}, an astronaut in orbit or a sailor out at sea).
  We discuss our findings which include the advantages of embedded visualised information in terms of precision compared to situated projected information with the accompanying limitations in addition to future improvements to our proposition.
  We conclude with the prospects of our work, notably the continuation and possibility of evaluating our proposition in a less controlled and real context in collaboration with our national space agency.
\end{abstract}



\begin{teaserfigure}
  \includegraphics[width=\textwidth]{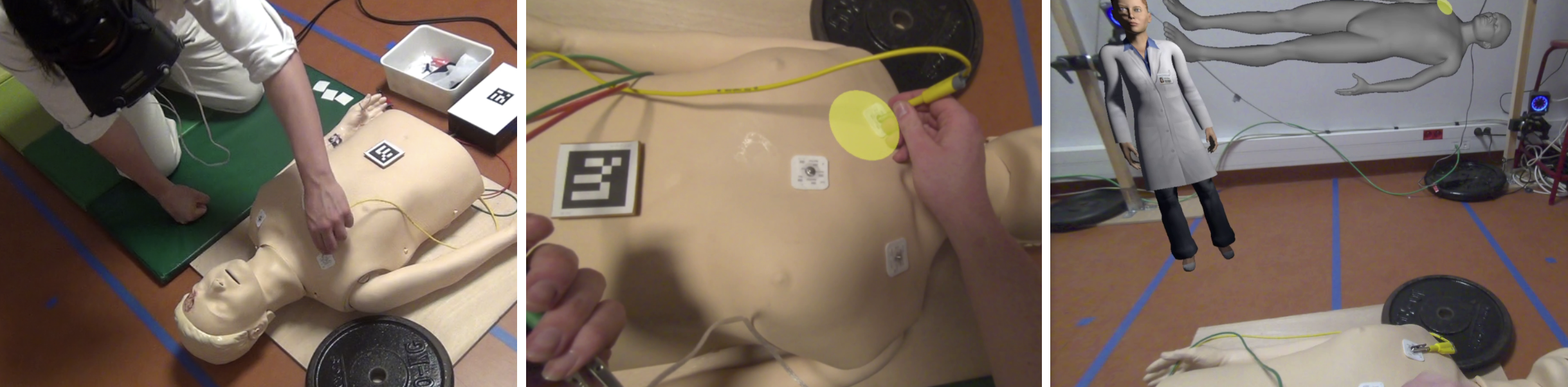}
  \caption{Screen captures of recordings during the experiment of the participant connecting the yellow cable to an electrode. From left to right: External view - Participant's view of the embedded visualisation instruction - Participant's view of the situated projected visualisation instruction}
  \Description{This is an image containing three screen capture of different perspectives including the two variants of our study. The first is an external view of a participant with an mixed reality headset performing an action on a manikin in front of them. The second is an internal view of what the participant sees during the embedded procedure. They currently see a yellow circle placed on top of an electrode and they are attaching a yellow cable on to this electrode. The third and final image shows the internal view of what the participant sees during the situated projected procedure. They are looking at a projected manikin with the same yellow circle as the previous image placed in the same location but on the projected manikin.}
  \label{fig:teaser}
\end{teaserfigure}


\maketitle

\section{Introduction}
\label{introduction}

In current research, a relevant term in Augmented Reality (AR) applications is that of \textit{situated visualisation}~\cite{white2009sitelens,willett2016embedded}, which places virtual objects in relation to their environment.
However the question of how to best display information using this type of visualisation still stays an important question to enable users to efficiently observe and treat potentially complex issues/tasks according to the context of use by using this emerging technology.

To find an answer to this, the field of \textit{Immersive Analytics}~\cite{marriott2018immersive}(IA) becomes of great importance as it is a field of study centered around using virtual displays to immerse the user in information placed in the surrounding space.
Since IA involves immersing the user by placing virtual information around them, the idea of integrating this information into the physical world around them becomes very fitting.
The combination of situated visualisation and IA is called \textit{Situated Analytics}~\cite{elsayed2015situated}(SitA). 
Defined by Thomas \textit{et al.}~\cite{thomas2018situated} as "the use of data representations organized in relation to relevant objects, places, and persons in the physical world for the purpose of understanding, sensemaking, and decisionmaking", SitA integrates IA into the physical environment of the user. 
According to this definition, SitA can play a large role in AR guidance, increasing the operator's spatial cognition of the situation, allowing for potentially more precise actions to be performed~\cite{wang2021role}. More precision allows for guidance through potentially more complex tasks and could benefit many fields greatly.
These complex procedures include ones such as medical diagnosis and surgery or even industrial maintenance and assembly. 
SitA techniques also allow for information to be displayed using AR with a diverse selection of visual design patterns~\cite{lee2023design}, and notably: \textit{embedded information}~\cite{willett2016embedded} and \textit{situated information}~\cite{white2009sitelens}. 

In this paper, we contribute to this research topic by studying and comparing these two design patterns in fields benefiting from different visual aids during complex procedures.
The context studied during this article is that of a medical emergency in a remote and isolated location in which we intend to provide AR instructions to the user.
On one hand, due to the partnership with our local space agency, the remote and isolated location studied is that of an astronaut carrying out basic life support on a crew member. Currently these procedures are primarily available on paper, traditional screens or even for the case of MobiPV~\cite{boyd2016mobipv} a wearable pair of smart glasses~\cite{HOPPENBROUWERS2017255}.
On the other hand, AR instructions need to be relayed efficiently and effectively so that the isolated user can employ their novice medical knowledge, expanded on by the information of an expert provided by our system.
For that, we compare embedded visualisation which makes use of \textit{physical data referents} to integrate virtual elements into the physical world and a situated projected visualisation which places virtual information relevant to it's environment yet not coinciding with it. 

Concretely, this article leads to the following contributions: 1) A scenario based evaluation of medical AR guidance using situated and embedded visualisation, 2) The design of two forms of visualisation for this context (embedded and situated projected) and 3) A user experiment comparing the two designs within our scenario.
This paper provides an overview of the developments in situated visualisation and AR guidance to set the context for this study. Next, we describe our system with the reasoning behind the technology we used. Then, we cover the design of our experiment, followed by the results and findings. We finish with a discussion and conclusion of the whole.

\section{Related Work}
\label{sota}

In this section, we present the main fields in which we aim to contribute to. 
Firstly, we address the field of situated visualisation and SitA, then the field of AR guidance.
After presenting the related work and concepts relevant to each of these topics, we position ourselves in respect to the scientific challenges in these fields.

\subsection{Situated Visualisation}
\label{situated_visualisation}

\textit{Situated visualisation} is a form of visualisation which extends the physical world with virtual objects or a "\textit{visualisation that is related to and displayed in its environment}" as defined by White \& Freiner~\cite{white2009sitelens}. White later defined the key characteristics common to all situated visualisations~\cite{white2009interaction}: 
1) Data in the Visualization is related to the physical context.
2) Visualization is based on the relevance of the data to the physical context.
3) Display and presentation of the visualization is in the physical context.

This definition was later expanded on by Willett ~\textit{et al.}~\cite{willett2016embedded} with the inclusion of \textit{physical data referents} upon which virtual objects are associated with. This inclusion brought another form of visualisation to the table called "\textit{embedded data representations}" which differs from situated representations by strictly coinciding spatially with the associated data referents as opposed to simply being in proximity to the data referents. 
The main benefit of situated and embedded visualisations is to allow a user to understand data \textit{in situ}, without splitting attention across physical and virtual objects.

In a survey by Bressa~\textit{et al.}~\cite{bressa2021s},
these two definitions for situated visualisation are stated to be the most prevalent definitions of situated visualisations in literature. 
Furthermore, the authors also mention the relevance of \textit{Situated Analytics}~\cite{elsayed2015situated}(SitA) which is used to encompass the fields of situated visualisation and \textit{Immersive Analytics}(IA). Both of these fields highly benefit from each other with IA being "\textit{the use of engaging, embodied analysis tools to support data understanding and decision making.}"~\cite{marriott2018immersive}.

\subsection{Situated Analytics}

A recent work expanding on the theme of SitA is the work done by Lee ~\textit{et al.}~\cite{lee2023design} which has categorised the various design patterns of situated visualisation based on other existing works and serves as a clear base to start from within the field. The work also describes many guidelines and constraints imposed on the application of situated visualisation.

Concerning the potential problems with designing a system with SitA, it must be noted that the efficiency of the situated information is highly dependent on the \textit{Extent of World Knowledge} (EWK), as stated by the lesser-known part of Milgram \& Kishino's work~\cite{milgram1994taxonomy}. 
With high EWK, the visualisations are strongly grounded to the real world whereas with low EWK, interactions become more and more necessary to counterbalance the lack of immersion and ability to extract information from it. 
Lee ~\textit{et al.}~\cite{lee2023design} also state that within the EWK, important factors include: \textit{referent density}, where too little makes tasks trivial and too many lead to readability and field of view issues; \textit{referent size}, which can lead to occlusion and impracticality issues, especially for embedded visualisations; and \textit{location awareness}, which is very dependent on the location and frequency of updates of the data referents.

With work still being done on the guidelines and categorisation in the field of situated visualisation as shown by Lee ~\textit{et al.}~\cite{lee2023design}, there still exists numerous scientific hurdles to tackle such as a solid foundational knowledge and taxonomy encompassing the whole topic (mentioned by Lee \textit{et al.} and other recent surveys~\cite{bressa2021s,satriadi2023proxsituated,shin2023reality}).
These newly defined taxonomies also need accompanying performance analysis to demonstrate their usage of which we contribute to with this article.

With the context of SitA addressed, a field which can highly benefit from it, is that of AR guidance. AR guidance aims to carry a user through a series of physical tasks, so having a way to direct the user's attention using visual indicators can be seen as a huge benefit which we will detail now.

\subsection{AR Guidance}
\label{ar_guidance}

AR guidance can be defined by \textit{AR instructions} which instructs the user about elements involved in the physical task, to help form the spatial cognition of the interaction. This can be any information related to the physical task which can be presented virtually such as the spatial relationship, operation method, etc.~\cite{wang2021role}.

A topic which is commonly studied by researchers is the way to effectively transfer sufficient knowledge to a novice user of an AR guidance system~\cite{baxter2012human,wang2021role}. The main difference between novices and experts is their mental image of the task~\cite{baxter2012human} and this means that, depending on the user's proficiency, there is a varying need for more spatial cognition when they are presented with a new task which needs to be performed effectively~\cite{wang2021role}. This spatial cognition plays a large role in the effectiveness of the AR guidance as it involves the cognitive ability of the user to solve spatial problems.
With this limit on the cognitive ability being different for all users, AR guidance systems cannot currently support high precision tasks~\cite{wang2021role} and needs to be synchronized with the user's cognitive needs to reduce their difficulty in cognition.
Wang ~\textit{et al.}~\cite{wang2021role} therefore designed their own system called UcAI (user-centered AR instruction) which has the goal to let a novice user play the role of the "\textit{thinker}" using information provided by an expert.

In literature, the two main forms of visual AR instruction type are \textit{Static} which reflects the physical state of an assembly object and \textit{Dynamic/Moving} which reflects the guiding method of an assembly object~\cite{feiner1993knowledge,wang2016multi,wang2022comprehensive,maffei2023dynamic,wang2021role}. 
The difference between these forms can be seen tested in the field of teaching, with an AR learning application~\cite{montoya2016evaluating} which states that dynamic content was found to be more helpful in understanding concepts by students. 

It is however important to note that when tasks become complex, dynamic information can also be distracting~\cite{wang2022comprehensive} and that concise visual information is not always sufficient to convey very complex information, making operation intention difficult to understand and leads to higher cognitive load for the user~\cite{mizell2001fundamentals,fiorentino2009tangible}.
The act of shifting the user's attention when providing information is also a cause for higher cognitive load as the user must remember the information and return to perform the action, which is especially important in fields such as industrial assembly~\cite{vanneste2020cognitive}.
On the other hand, simplified information can lead to the aforementioned disparity between novice and expert users~\cite{wang2022comprehensive}.
It is relevant to mention that the design of AR guidance systems highly benefits from multi-modal interactions to trigger visual elements using tools such as head and eye tracking, haptic feedback and affective computing~\cite{wang2022comprehensive} which can compensate for simplified information. These interactions generally solve the problems of occlusion and visual clutter in AR~\cite{truong2021user} but can also lead to higher completion times and possible overload which is the case for the auditory interactions in~\cite{marquardt2020comparing}. 

An interesting example of AR guidance in terms of task model is that of the Holopit project~\cite{lallai2021engineering} which employs various situated visualisation guidance techniques to guide a novice pilot through a series of flight procedures using the knowledge and instructions provided by an expert.
A downside with their system was that the expert had to place the virtual objects over their respective locations which highlights an issue with the identification of data referents.

The various applications of XR are almost limitless as the fields in which virtual additions benefit the user is down to imagination. 
The main hold-back however is the existence of social and technological barriers for entry into the various fields of application~\cite{fast2018testing}.
Among the domains of application that have seen the appearance of SitA and AR guidance, the medical field is one that is of great interest to researchers due to the often complex and non-deterministic tasks such as surgery and diagnosis.

\subsection{AR Guidance for Medical Assistance}
\label{medical_ar}

The idea of using Extended Reality (XR) for surgical simulation was proposed by Satava's paper~\cite{satava1993virtual}, which used an off-the-shelve HMD in addition to a DataGlove to interact with a virtual abdomen to provide a virtual reality training simulator.
As for AR medical applications, the early thoughts were summarized through the review of Tang~\textit{et al.}~\cite{tang1998augmented}.
Yet, even today, many papers show promising results for medical application but little literature or actual surgical application prove its real world usage outside of training and education~\cite{barsom2016systematic,chen2017recent,eckert2019augmented,tang2020augmented}.

Nevertheless, the appearance of the MS Hololens made a huge mark in the field, being used in numerous medical prototype/proof-of-concept studies as mentioned in the review from Park~\textit{et al.}~\cite{park2021review}.
Recent applications of AR guidance include projects such as AR-Coach~\cite{ebnali2022ar} which simulates medical assistance with direct contact with the AR avatar of an expert in a control center.
Other research has proposed evaluations on the need for basic life support in out-of-hospital cases for example sudden cardiac arrest~\cite{9978596,info:doi/10.2196/14910} demonstrated through medical training simulations such as Viewpoint~\cite{info:doi/10.2196/28595} and RescuAR~\cite{javaheri2023rescuar}.

With our work, we contribute to the AR guidance and situated visualisation fields applied to a medical context by proposing an evaluation of the impact of visual AR guidance with the addition of SitA all whilst taking into account the mental image of an expert in order to improve the actions of a novice similar to the works of Baxter~\cite{baxter2012human} and Wang~\textit{et al.}~\cite{wang2021role}.

\section{Design of our System}
\label{system_design}

\subsection{Context}
\label{context}

Our work is in line with the objective of collaboration between two users, one distant from the other (one on earth whilst the other is on Mars or the ISS for example) or even a team in an isolated area in need of the assistance of an expert (space, desert or even at sea).
In such situations, isolation may be not simply geographical but even temporal (communication latency or even failure).
These environments call for appropriate technical, technological, design and formalisation choices.
We therefore explain our choices which we made in the following sections.

It should be noted that these choices have been based on another study carried out in parallel with this work. This study is about a different subject and its results in no way influence the results of this current work. 
Within the context of this paper, we are not concerned with the collaboration aspect between two users. We are only concerned with the potentially distant user and their isolated situation and therefore the information displayed using AR.
We also state that the system was designed with the fact that it should also function correctly in various isolated situations such as at sea or within a weightless environment as per the requirements of our local space agency.

\subsection{Technology}
\label{technology}

We used the Varjo XR-3 VST HMD~\cite{varjo} for AR visualisations and interactions (see Figure~\ref{fig:teaser}, left picture). We chose this HMD due to the high resolution 4K cameras of the video which makes the user more able to perform tasks with higher detail visuals as well as features such as eye tracking (used to track where the user's attention is throughout the procedure and will be present in the results of this work) and hand-tracking using ultra-leap depth sensor cameras.
Another requirement when designing our system was to not use the internal position tracking system integrated in the Varjo HMD nor the classical way with VR Base Stations. 
These technical requirements are imposed by the different types of isolated environments (space, desert, open sea) mentioned in section~\ref{context}. 
These isolated environments do not necessarily allow us to make use the HMD's incorporated inertial sensors.
In this respect, the use of infra-red (IR) cameras and motion capture technology using marker constellations was concluded as the most appropriate for the different types of aforementioned isolated situations.
We therefore used the Optitrack 13 Prime~\textsuperscript{x} W cameras and Motive~\cite{optitrack} to get the tracking data.
We integrated into our IR camera tracking system, a hybrid system using visual markers (QR codes) picked up by the HMD's cameras for virtual element positioning.
This hybrid system allows us to be more precise on the positioning of virtual elements on top of physical ones. When the user needs a close-up view, the visual markers take priority over the IR tracking. 
This allows our system to compensate for possible disturbances of our IR tracking such as lighting or obscuring the reflective tracking surfaces.
These technological choices are motivated from three separate weightless flight campaigns as well as one at sea to prove the robustness in those environments. However, the results of these experiments are not treated in this article.
Additionally, we make use of a physical manikin (Laerdal Resusci Anne) in our experiment.
This manikin plays the role of the patient with an acute traumatic abdominal pain and is equipped with a QR code for our visual tracking.

\subsection{Design of the Program}
\label{program_design}

The conception of our virtual environment is based in MASCARET~\cite{chevaillier2012semantic,saunier2016methodology} which is a metamodel based on UML/SYSML which allows experts to describe the semantics of the environment, behaviours of components of the environment as well as human activities.
In our case, the model is written by medical experts who fill out our environment with the necessary medical equipment such as a medical patient monitor.
This equipment is described using class diagrams and attributes (such as heart rate, oxygen saturation levels, \textit{etc.}).
This can also include features such as highlights over specific areas.
For complex and autonomous behaviours, state machines are used. 
Domain specific procedures (in our case medical procedures) are defined by UML activities which organise actions involving various resources from the environment. 
Actions are then executed by the agents of our scene who play specific roles.
These agents are either human such as the person using our system in our case or virtual agents such a medical guide (see Figure~\ref{fig:teaser}, right picture).

In the environment relevant to our study, we have 4 main virtual elements (See Figure~\ref{fig:unity_view}):

\begin{figure}[!h]
  \centering
  \includegraphics[width=0.6\linewidth]{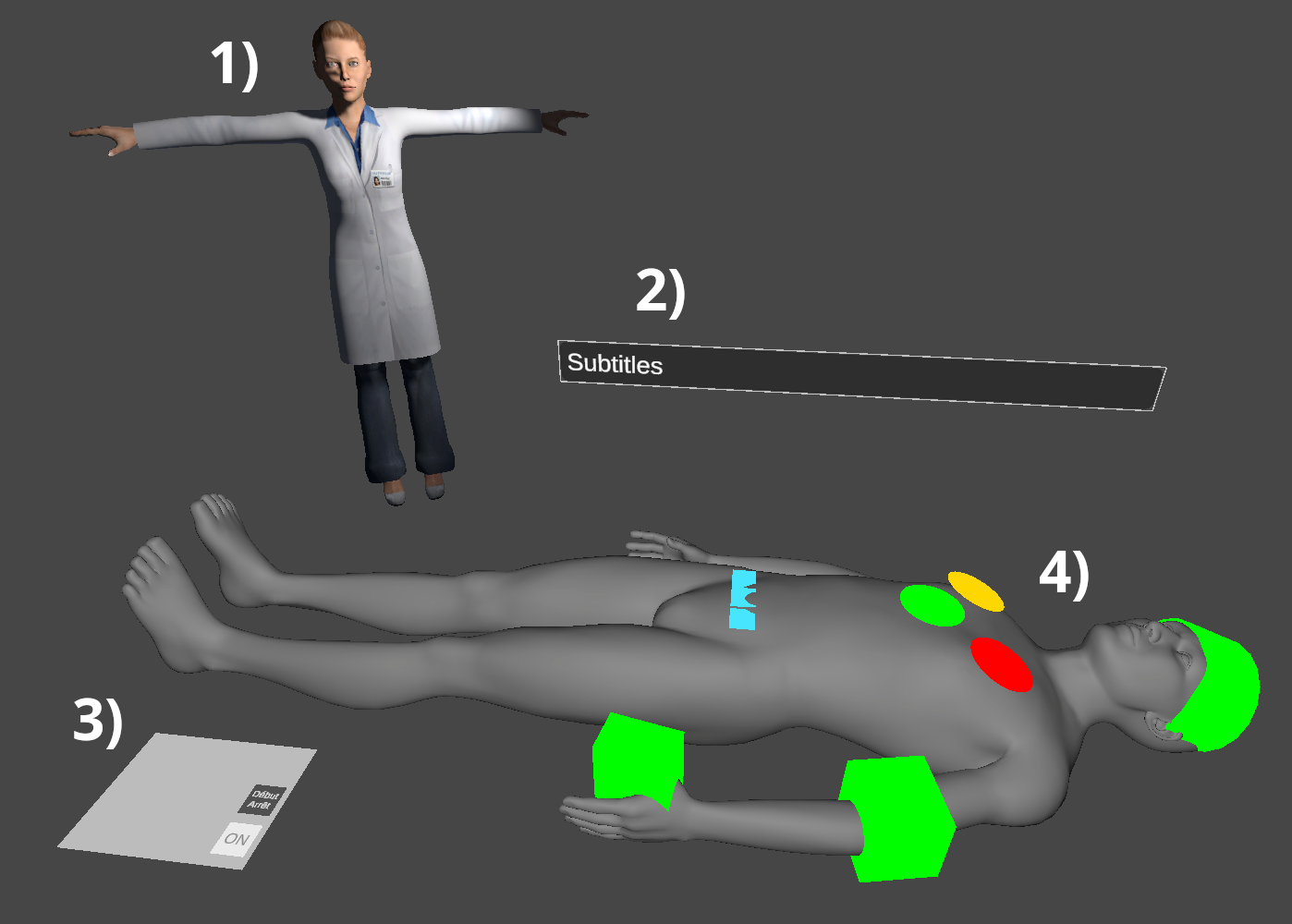}
  \caption{Virtual elements: ECA (1); Subtitles (2); Patient Monitor (3); Virtual manikin \& virtual indicators (4)}
  \Description{Screen capture of the virtual objects (An ECA or Embodied Conversation Agent, a subtitle bar containing the contents of the intructions spoken by the ECA, a patient monitor showing the vitals of the patient and the highlighted areas of interest represented on a virtual representation of the patient in unity}
  \label{fig:unity_view}
\end{figure}

\begin{figure}[!h]
  \centering
  \includegraphics[width=\linewidth]{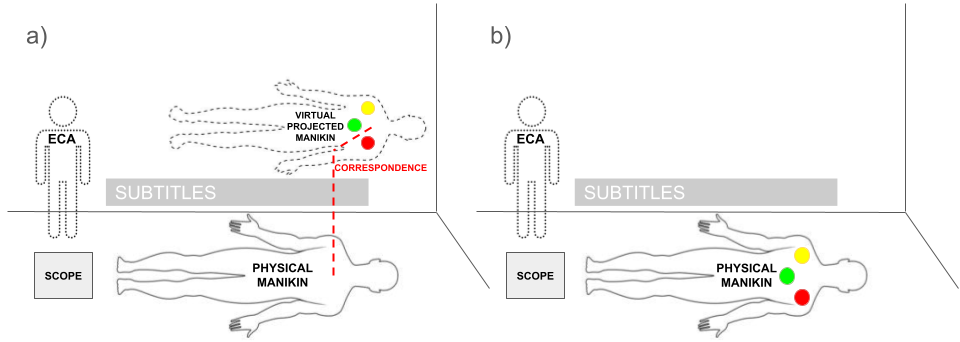}
  \caption{Comparative diagram of the situated projected visualisation environment (a) and embedded visualisation environment (b)}
  \Description{Comparison between both visualisations. The situated projected visualisation shows the virtual manikin projected above the physical one and the areas of interest are placed on the virtual manikin. The embedded visualisation does not contain a virtual manikin and the areas of interest are placed on the physical manikin.}
  \label{fig:visualisation_comparison}
\end{figure}

\begin{enumerate}
    \item \textit{An Embodied Conversational Agent} (ECA). As MASCARET implements SAIBA architecture~\cite{vilhjalmsson2007behavior}, this allows the implementation of an Embodied Conversational Agent (ECA) into our environment. This ECA is able to relay the information linked to each task orally with occasional animations to show the necessary action to perform in the same manner as proposed by Collins~\textit{et al.}~\cite{collins2019modelling}. The ECA's job is to assist (at least with its presence) and relay information and instructions about the procedure (in this case medical, hence the choice of a doctor in a white coat).
    \item \textit{Subtitles}. Subtitles appear and update when the ECA is conveying instructions to give the operator a chance to read the instructions as well. They fade once the ECA has finished reading but reappear if the operator wishes to re-listen to the ECA.
    \item \textit{Patient Monitor}. A tool which was used during the medical instructions for reading values such as the heart rate or oxygen levels. In the context of this experiment, this device is simply represented by visual element placed in the world on a physical white box. The visual values displayed on the Patient Monitor change depending on the task progression. Outside of this experiment, a real physical Patient Monitor could be used with the virtual information superposed on its screen.
    \item \textit{Areas of Interest}. Concerning the first of our two variations of visualisation, our situated projected visualisation is presented as a virtual reflection or copy of the physical manikin, upon which virtual indicators will appear and disappear depending on the tasks throughout the procedure as can be seen in Figure~\ref{fig:visualisation_comparison}~(a).
\end{enumerate}

In this visualisation format, the user must create a mental correspondence between the physical areas (ex. right shoulder of the physical manikin) and the virtual areas (ex. area of interest on the right shoulder of the virtual manikin).
It must be noted that the position of this virtual manikin is always placed statically above and aligned with the physical manikin and opposite the area in which the operator performs their tasks.

On the other hand, the embedded visualisation places the virtual indicators directly on the manikin instead of a virtual projection, overlapping the physical position where the action or placement needs to be performed which can be seen in Figure~\ref{fig:visualisation_comparison}~(b).

In correspondence with Lee \textit{et al.}'s work (read \S\ref{situated_visualisation}), our system is respectively composed by the following patterns (see Figure~\ref{fig:unity_view}):
\begin{description}
    \item[1.] Embodied Conversational Agent: Ghost pattern
    \item[2.] Subtitles: Panel pattern
    \item[3.] Patient Monitor: Panel pattern
    \item[4.a.] Our "Embedded" information: Glyph or Decal pattern
    \item[4.b.] Our "Situated Projected" information: Mix of Proxy and Virtual Mirror
\end{description}

We make use of the Panel pattern to display information as does the typical display to make use of their familiarity and intuitivity as do a vast majority of AR projects (For example ScalAR which uses semantic-based AR design to place Panels~\cite{qian2022scalar}). 
The Ghost pattern is used for the ECA to complement the scene to add further context to actions (such as our animations and avatar to complement the medical context) such as the "The Invisible Train" which made a game naturally more intuitive to users who might not have prior AR experience~\cite{wagner2005towards}.
These two types of patterns are stated by Lee \textit{et al.} to be the most common types for AR instruction (Panels followed by Ghosts) mainly due to their simplicity, familiarity and easy implementation into projects.
In the same way as Lee ~\textit{et al.}~\cite{lee2023design}, we are also studying the visual element of the system and not the interactions. We do however note of the importance of interactions in sense-making as do Lee ~\textit{et al.} including the relevance that comes from AR guidance~\cite{truong2021user}. These elements are further elaborated in Section~\ref{conclusion} in terms of prospectives.

\section{Experiment}
\label{experiment}

Our experiment has the goal to measure the impact of two forms of visual presentation of information (situated projected visualisation and embedded visualisation) on the operator's performance during a simulated medical procedure.

The procedure which was carried out by the operator is that of a medical emergency centered around a manikin, placed in front of them. The operator was equipped with the AR hardware as well as the relevant medical equipment needed to fulfill the various tasks.
This procedure was written and proposed by a doctor specialised in medical user autonomy in isolated areas in space and corresponds to a procedure called \textit{"diagnosis of acute traumatic abdominal pain"}.

\subsection{Tasks}
\label{tasks}

As mentioned previously, our experiment is based around performing a medical procedure. This procedure is divided into 23 individual tasks which remain indifferent for all operators with the exception of our two visualisation formats.
Tasks were classified into three categories to assess operators' performance in speed, precision and attention.
The tasks types are as follows: 
\begin{itemize}
    \item CONSULTATION: Requires the operator to consult and report visible information on the patient or instruments. These tasks test the operator's ability to extract details from displayed information.
    \item ACTION: Involves the operator to perform a physical action, such as pressing a button or connecting cables. These tasks evaluate the ease of which instructions are followed and performed.
    \item PRECISION: Requires more delicate actions, such as precise placement of electrodes or feeling precise areas on the manikin. These tasks measure the precision and accuracy of the operator's interactions with the environment.
\end{itemize}
The various tasks performed during the test protocol are as stated in Table~\ref{tab:task_table}. This table presents each of the 23 tasks which constitute our medical procedure. 
This table contains the task number, name, action category, details and finally the conditions which we defined to determine whether or not the task was correctly performed.

{
\tiny
\begin{table*}
  \caption{Tasks to be performed by the operator during the test protocol}
  \label{tab:task_table}
  \begin{tabular}{p{0.8cm} p{1.5cm} p{1.5cm} p{6.2cm} p{6cm}}
    \toprule
    \centering Task Number & \centering Task Name & \centering Task Category & Task Description & Success Criteria\\
    \midrule
    \centering 1/23 & \centering Check for vomit & \centering CONSULTATION & Check the patient for vomit and report vocally the result & The area of interest needs to be observed by the participant \\
    \centering 2/23 & \centering Check for diarrhea & \centering CONSULTATION & Check the patient for diarrhea and report vocally the result & The area of interest needs to be observed by the participant \\
    \centering 3/23 & \centering Check for pain & \centering CONSULTATION & Check the patient for pain and report vocally the result & The area of interest needs to be observed by the participant \\
    \centering 4/23 & \centering Turn on monitor & \centering ACTION & Turn on the patient monitor by pressing the ON button & Finger must be on top of virtual element on the box \\
    \centering 5/23 & \centering Place Green Electrode & \centering PRECISION & Place an electrode patch at the location of the heart & The electrode must be positioned within the indicated zone without overlapping the limits, with a tolerance of +/- 1 cm \\
    \centering 6/23 & \centering Place Red Electrode & \centering PRECISION & Place an electrode patch on the left shoulder & The electrode must be positioned within the indicated zone without overlapping the limits, with a tolerance of +/- 1 cm \\
    \centering 7/23 & \centering Place Yellow Electrode & \centering PRECISION & Place an electrode patch on the right shoulder & The electrode must be positioned within the indicated zone without overlapping the limits, with a tolerance of +/- 1 cm \\
    \centering 8/23 & \centering Connect Yellow Electrode & \centering ACTION & Connect the yellow cable to the right shoulder electrode & The cable must be connected to the correct electrode \\
    \centering 9/23 & \centering Connect Red Electrode & \centering ACTION & Connect the red cable to the electrode on the left shoulder & The cable must be connected to the correct electrode \\
    \centering 10/23 & \centering Connect Green Electrode & \centering ACTION & Connect the green cable to the electrode at the location of the heart & The cable must be connected to the correct electrode \\
    \centering 11/23 & \centering Attach Tension Cuff & \centering ACTION & Place the tension cuff on the patient's arm & Cuff must be on the correct arm, must not be inside out and put on the bicep of the manikin \\
    \centering 12/23 & \centering Set Blood Pressure & \centering ACTION & Press the PNI icon on the patient monitor screen & Finger must be on top of virtual element on the box \\
    \centering 13/23 & \centering Start Blood Pressure  & \centering ACTION & Start blood pressure measurement by pressing the start/stop button & Finger must be on top of virtual element on the box \\
    \centering 14/23 & \centering Place SPO2 & \centering ACTION & Pinch the SPO2 sensor on one of the patient's fingers & Sensor must be put on a finger on the left hand \\
    \centering 15/23 & \centering Measure Heart Rate & \centering CONSULTATION & Announce the heart rate (HR) value, displayed in green on the monitor & The correct number(s) must be said aloud \\
    \centering 16/23 & \centering Measure Blood Pressure & \centering CONSULTATION & Announce the blood pressure value, displayed in orange on the monitor & The correct number(s) must be said aloud \\
    \centering 17/23 & \centering Measure Oxygen Saturation & \centering CONSULTATION & Announce the SPO2 value, displayed in blue on the monitor & The correct number(s) must be said aloud \\
    \centering 18/23 & \centering Palpate Q1 & \centering PRECISION & Palpate the left iliac area and notify the nurse if the area is soft or hard & The hand gesture must be within the same axis as the visual representation along the length of the manikin \\
    \centering 19/23 & \centering Palpate Q2 & \centering PRECISION & Palpate the suprapubic area and notify the nurse if the area is soft or hard & The hand gesture must be within the same axis as the visual representation along the length of the manikin \\
    \centering 20/23 & \centering Palpate Q3 & \centering PRECISION & Palpate the right illiac area and notify the nurse if the area is soft or hard & The hand gesture must be within the same axis as the visual representation along the length of the manikin \\
    \centering 21/23 & \centering Set Up Infusion & \centering ACTION & Connect the tubing to the infusion catheter and screw it on & The infusion must be attached and screwed on to the attachment on the arm \\
    \centering 22/23 & \centering Heart massage & \centering ACTION & Place your hands at chest level and apply pressure four times & 4 massages must be performed with enough pressure and in the correct location \\
    \centering 23/23 & \centering Wrap Bandage & \centering ACTION & Wrap the bandage around the patient's head & The bandage must cover the injury on the manikin's head \\ 
  \bottomrule
\end{tabular}
\end{table*}
}

\subsection{Participants}
\label{participants}

We recruited a total of 27 workplace safety/first-aid qualified following EN ISO 14971 regulation participants (8F, 19M), with 14 using the "Embedded" visualisation (3F, 11M) and 13 using the "Situated Projected" visualisation (5F, 8M). 
This amount of participants was chosen to assure that our quantitative values were sufficient enough to perform the chosen tests whilst keeping it to a feasible amount to conduct the experiment.
Participants were also assigned their visualisation format randomly to avoid selection bias.
Overall, the data from 4 participants (3M in the "Embedded" group and 1F in the "Situated Projected" group) was excluded due to system-related issues such as the malfunction of their eye tracking which compromised the validity of their data.
The selected participants for the experiment are all holders of a valid workplace safety/first-aid certification respecting EN ISO 14971 regulation. This assures that the participants have a solid foundation of base-level first-aid knowledge.
This type of participant profile is very similar to that of our distant operator in an isolated area such as an astronaut in orbit, a sailor out at sea or perhaps a soldier in hostile territory. 
In these situations, the operators involved are not necessarily experts in medicine. 
Although trained in the latter, they do not practise it on a daily basis, and their knowledge may become "latent".
This type of profile, suggested by our specialist collaborators in space and remote medicine, allowed us to check whether our guidance system served as a suitable resource for retrieving latent knowledge as mentioned in section~\ref{ar_guidance}. Finally, the ethics committee of the authors’ institution has approved this experimental protocol (agreement CER AUB n°2406204). 

\subsection{Experimental design and procedure}
\label{experimental_design}

The participants were to carry out the tasks in a procedural fashion with either a \textit{situated projected} or \textit{embedded} format of visual presentation. 
Both variants of visualisation were not explained to the participants, they were only presented with their pre-assigned one.

This experiment consisted of four distinct phases: 

\subsubsection{Presentation phase}
\label{presentation_phase}
The objective of this phase was to present the experiment to the participant so that they have the minimum amount of knowledge to perform the various tasks, excluding their novice medical knowledge and information about the different types of visualisation.
Upon arrival, we described to them that they will be performing a medical procedure using our system.
They were informed about the two available keywords which the participant could state aloud during the procedure.
We explained that the keyword "NEXT" was used to move on to the next task once they had deemed their current task complete. 
The other keyword told to them was "REPEAT" which makes our ECA repeat the instruction of the current task and have the subtitles reappear for them.
The medical equipment was also briefly presented to avoid any confusion towards the naming of any of the medical instruments in question. 
The whole took around 5 minutes to complete.

\subsubsection{Familiarisation phase}
\label{familiarisation_phase}
A "familiarisation" phase was carried out before the test procedure. This phase was designed to allow the participants to get to know the system and to briefly get used to the AR experience as well as the basic types of actions which they would need to perform before continuing with the full test procedure.
They had to perform a CONSULTATION task followed by an ACTION task to show an action requiring verbal feedback and the other involving one of the two formats of presenting information.
This phase took around 2 minutes to complete.

\subsubsection{Test procedure phase}
\label{test_procedure_phase}

For the main test procedure phase, the participants needed to complete all of the procedure (\textit{i.e.}, the 23 different tasks) so that we could evaluate our two types of visualisation.
This was done by our collection of quantitative data during this phase through our program. The time to complete each task, the number of repetitions requested and the location and time spent looking at each location of the eye-tracking of each participant was measured. The precision and correctness of each action was monitored visually throughout the test procedure and with additional recordings and images taken to assure consistency.
The participants had to complete each task one by one unbeknownst of the following instructions without the option to return to a previous task.
This procedure took the participants on average 6 minutes to complete.

\subsubsection{Post-experiment procedure phase}
\label{post_experiment_phase}
Once the test procedure was completed, we collected some more data based on the workload and usability of our system as well as some qualitative data both to evaluate the perceived workload and for general feedback on the experience. 
To fulfill these requirements, the participants were therefore requested to fill out a System Usability Scale~\cite{brooke1996sus} and NASA Task Load Index~\cite{hart2006nasa}.
An interview was also carried out after the questionnaires to acquire the feedback and opinions of the system which we needed but also to back up some of the results from the questionnaires.
This phase took around 10 minutes to complete.

\subsection{Hypotheses}
\label{hypotheses}


We hypothesised that the embedded format of visualisation will be more efficient than the situated projected format in task completion (H\textsubscript{1}) as it would lead to the operator performing less overall shifts in the user's attention and less head movements. Using physical data referents would increase the operator's EWK inducing a more fluid and direct retrieval of tools and placement of objects making tasks easier to complete~\cite{milgram1994taxonomy,lee2023design}. 
This would also mean that the user's spatial cognition would be higher and cognitive load to be lower~\cite{wang2021role}.
We also suspected that embedded information will be more appropriate compared to situated projected information when performing actions involving a higher precision (H\textsubscript{2}). The operator should have the relevant locations highlighted, which we believe will make it more intuitive for them and hence lead to more precision. This highlighting of areas on the manikin should improve the users mental image and therefore allow them to understand more complex tasks such as precision tasks~\cite{baxter2012human}.
Finally we believe that the users will see less of a cognitive load for embedded visualisations than situated projected and will also be more willing to use embedded information (H\textsubscript{3}). This returns to some of our previous points (see Section~\ref{ar_guidance}) with the user shifting their attention more for situated projected information which should lead to higher cognitive load~\cite{vanneste2020cognitive,wang2021role}.
We therefore made the following hypotheses with these points in mind:
\begin{itemize}
    \item H\textsubscript{1}: Information presentation will have a significant impact on the efficiency of task completion, with embedded visualisations enabling a faster total completion time compared to situated projected visualisations.
    \item H\textsubscript{2}: The precision of actions will be improved with embedded visualisations compared to situated projected visualisations, especially for tasks requiring a higher precision.
    \item H\textsubscript{3}: Users will perceive a lower cognitive load and greater acceptability for embedded visualisations compared to situated projected visualisations.
\end{itemize}

\section{Results}
\label{results}

The statistical analysis tests carried out between the Embedded information group (E) and Situated Projected information group (SP) used a significance threshold for $p$ of $0.05$ and error bars with a confidence interval of $95\%$ are represented on all of the following bar charts. 
These tests were also performed between groups for each of the three action types to monitor precision and overall performance, attention and workload.

\subsection{Task Efficiency}
\label{efficiency}

To evaluate the efficiency of the task completion, we measured the total time taken for each task, number of times help was requested, number of requests to have the instruction repeated and the eye-tracking results in both time and gaze shifts per participant using either the embedded or situated projected visualisations.

\begin{figure}[!h]
    \centering
    \includegraphics[]{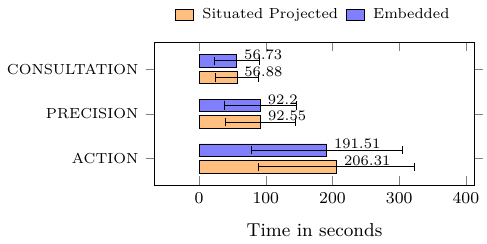}
    \caption{Bar chart showing average completion times for each task variety for both visualisation methods}
    \Description{Bar chart showing average completion times for each task variety for both visualisation methods. CONSULTATION took 56.73s for Embedded and 56.88s for Situated Projected on average; PRECISION took 92.2s for Embedded and 92.55s for Situated Projected on average; ACTION took 191.51s for Embedded and 206.31s for Situated Projected on average.}
    \label{fig:avgtime_chart}
\end{figure}

We applied Mann-Whitney U and Kruskal-Wallis tests due to the non-parametric nature of our data which does not follow normal distribution.
These tests were applied to the total execution times of each participant to complete all tasks as well as every task per category, keeping the independence necessary to perform each test.

Although the fastest completion times were 272.75s for E and 277.45s for SP, and the slowest times were 405.56s for E and 559.56s for SP, statistical analysis revealed no significant difference in overall task times ($p=0.926$), errors ($p=0.368$), nor help requests ($p=0.264$) between the groups ($p > 0.05$). 
This suggests that overall there is no evident effect on the completion time of the task depending on the format of visualising information.
This fact can be seen on Figure~\ref{fig:avgtime_chart} with ACTION type tasks proving only to be slower on average for SP information than E information.

We received similar results for each type of action with the number of times the participant requested for the instructions to be repeated ($p=0.971$).
This is mainly due to the fact that these features were not used often by the participants not even taking in to account the format of presenting the information.

On the other hand, the data from our eye-tracking data from the HMD showed some notable results.
The data however contained some parasitic values which were not uniform over both forms of visualisation due to the additional virtual projected patient leading to an overall higher number of results. 
These values were therefore filtered to obtain only the theoretically valid results. 
This was qualified by the time the operator spent looking at a virtual object. 
If the time was superior to 200ms, we deemed that the operator's attention was focused on that virtual object, otherwise the value was filtered. 
We found the same significant pattern of results for all types of action for gaze shifts which is in line with our expectations due to the existence of the potentially disruptive virtual patient.

Concerning the time spent looking at each virtual element, we also found some evidence towards there being a difference in time between visualisations for participants looking at the subtitles and the virtual patient ($U=120.0,p=0.001,r=0.693$). Additionally, more time was spent looking at the manikin for tasks in the "precision" category for the embedded format (E\textsubscript{average}=80.74s) than the situated projected format (SP\textsubscript{average}=60.45s) ($U=28.000,p=0.021,r=-0.488$).

\subsection{Action Precision}
\label{precision}

\begin{figure}[!h]
    \centering
    \includegraphics[width=.24\linewidth]{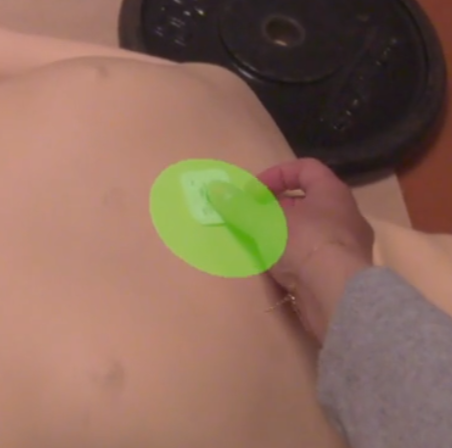}
    \includegraphics[width=.24\linewidth]{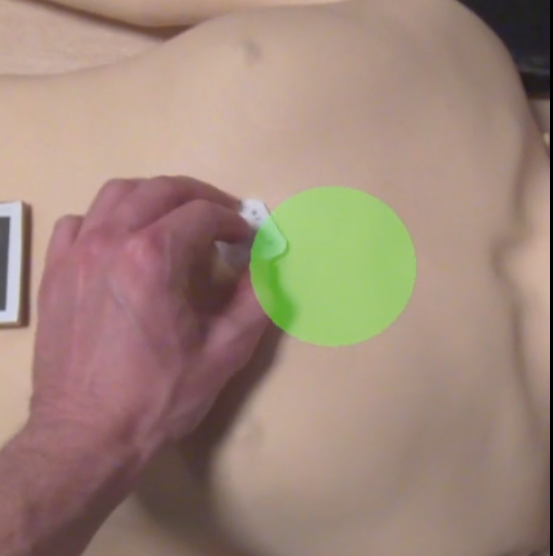}
    \includegraphics[width=.24\linewidth]{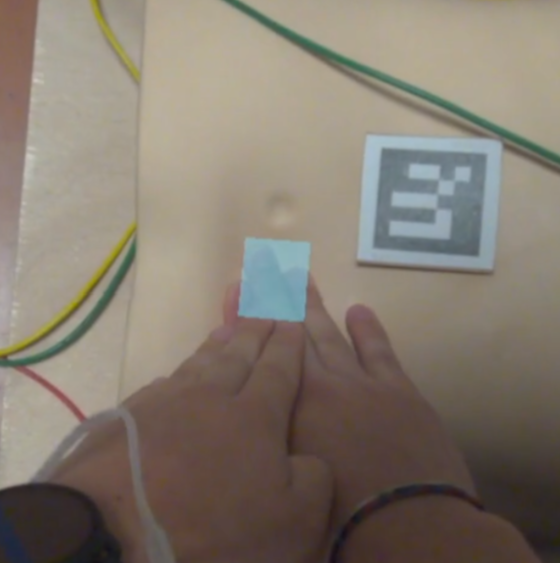}
    \includegraphics[width=.24\linewidth]{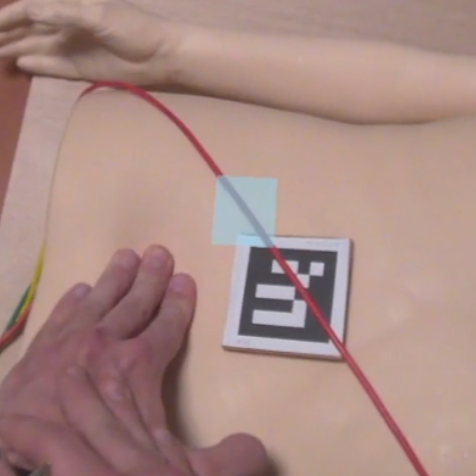}
    \caption{Valid and invalid placements for electrodes, then palpation tasks in embedded visualisation (left to right)}
    \label{fig:precision_check}
    \Description{4 images from left to right: The view of a participant placing an electrode within the green highlighted circle/area of interest on the manikin, a participant placing an electrode outside of the green highlighted circle/area of interest on the manikin, a participant performing a palpation within the blue highlighted square/area of interest on the manikin and a participant performing a palpation outside of the blue highlighted square/area of interest on the manikin}
\end{figure}

Precision was assessed by evaluating the placement of objects during tasks of the PRECISION category (Refer to Table~\ref{tab:task_table} for more details). 
Tasks involving the placement of electrodes were measured through the accuracy of the final placement of the electrode, comparing it to the respective visually highlighted position either directly on the manikin (for the embedded format) or on the projected virtual manikin (for the situated projected format). 
For the situated projected format, a post procedure comparison was performed using recordings and the perspective of how the participant observed the virtual manikin (according to the participant recorded view/video).
For palpation tasks, the area in which the operator placed their fingers was compared to the highlighted areas.
Figure~\ref{fig:precision_check} shows example placements for both tasks. 

\begin{figure}[!h]
    \centering
    \includegraphics[]{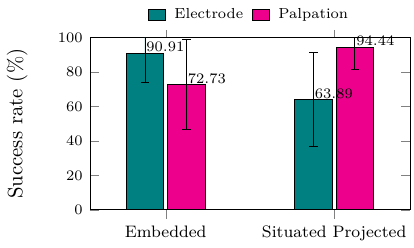}
    \caption{Column chart showing the success rate for each precision task for both visualisation methods}
    \label{fig:precision_chart}
    \Description{Column chart showing the success rate percentage for each precision task for both visualisation methods. For the palpations: 72.73\% for Embedded and 94.44\% for Situated Projected. For the electrodes: 90.91\% for Embedded and 63.89\% for Situated Projected}
\end{figure}

These comparisons were based on either the placement inside of the visual indication for E or based on an image comparison, using the same perspective and taken after the procedure was finished, for SP.
The results of both of these formats involved the human visual validation of each task using the collected data (recordings from two camera perspectives and photos) to state whether the task was performed within the indications or not. The whole analysis was performed twice each by a different person to assure the validity of the results.
We used Chi-squared tests as the data was boolean producing a 2x2 contingency table which allows us to follow up with a Fisher's Exact test.
Using these tests, we found some evidence that the E format was more accurate for electrode placement ($\chi^2=7.057, p=0.007$) whereas the SP format was more accurate for fingertip palpations ($\chi^2=6.060,p=0.013$).
This indicates that not all precision tasks benefit from a single type of visualisation which can be seen comparing the success rates of each task (see Figure~\ref{fig:precision_chart}).

Considering both types of precision action as a whole, the tests showed no evidence surely due to their opposite nature ($\chi^2=0.154,p=0.158$).
The same result was found when applying a Spearman's Rank Correlation test ($\rho=-0.199,p=0.364$).

\subsection{Perceived cognitive load and Usability}
\label{usability}

The usability of the system was assessed using the NASA-TLX and SUS questionnaires.
Mann–Whitney U tests were also used to compare the results with significant differences observed in the overall scores for both the NASA-TLX ($U=101.0, p=0.034, r=0.449$) and the SUS ($U=30.5, p=0.031, r=-0.456$) equally showing in Spearman's rank correlation tests for the NASA-TLX ($\rho=0.468,p=0.024$) and SUS ($\rho=-0.460, p=0.027$). 
Overall, the participants found that embedded visualisations lightened the workload through the statistical tests and the average scores (SUS: SP\textsubscript{average}=68.33, E\textsubscript{average}=81.59 ; NASA-TLX: SP\textsubscript{average}=37.67, E\textsubscript{average}=25.64).

\begin{figure}[!h]
    \centering
    \includegraphics[width=\linewidth]{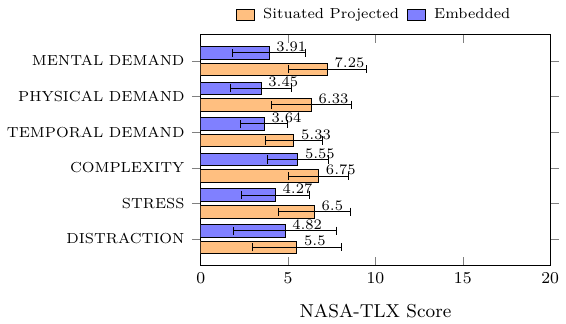}
    \caption{Bar chart showing the average results of the NASA-TLX questions for both visualisation methods}
    \label{fig:nasatlx_chart}
    \Description{Bar chart showing the average results of the NASA-TLX questions for both visualisation methods. The embedded format had 3.91 for MENTAL DEMAND, 3.45 for PHYSICAL DEMAND, 3.64 for TEMPORAL DEMAND, 5.55 for COMPLEXITY, 4.27 for STRESS and 4.82 for DISTRACTION. The situated projected format had 7.25 for MENTAL DEMAND, 6.33 for PHYSICAL DEMAND, 5.33 for TEMPORAL DEMAND, 6.75 for COMPLEXITY, 6.50 for STRESS and 5.5 for DISTRACTION.}
\end{figure}

Despite the individual questions in the NASA-TLX questionnaire being higher for situated projected than embedded as shown by Figure~\ref{fig:nasatlx_chart}, the statistical tests showed very little statistical evidence.

\begin{figure}[!h]
    \centering
    \includegraphics[width=\linewidth]{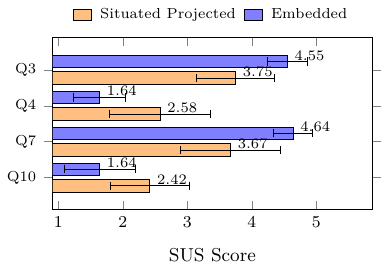}
    \caption{Bar chart showing the average results of the significant SUS questions for both visualisation methods}
  \label{fig:sus_chart}
  \Description{Bar chart showing the average results of the significant SUS questions for both visualisation methods. The embedded format had a score of 4.55 for Q3, 1.64 for Q4, 4.64 for Q7 and 1.64 for Q10. The situated projected format had a score of 3.75 for Q3, 2.58 for Q4, 3.67 for Q7 and 2.42 for Q10}
\end{figure}

This included some weak evidence towards the user's stress being higher for the situated projected format from the questionnaire ($U=34.500,p=0.055,r=-0.404$).
On the contrary, the SUS individually showed a difference for question 3 (\textit{"I think that this service is easy to use."}) ($U=33.5,p=0.029,r=-0.417$), indicating that the system overall was easier to use when embedded visualisations were employed (Q3: SP\textsubscript{average}=3.75, E\textsubscript{average}=4.55).
Additionally, some weak evidence was found for questions 4 (\textit{"I think I'll need the help of a technician to be able to use this service."}), 7 (\textit{"I imagine that most people would be able to learn to use this service very quickly."}) and 10 (\textit{"I need to learn a lot before I can use this service."}) ($Q4:U=94.0,p=0.075,r=0.359$;$Q7:U=39.5,p=0.082,r=-0.340$;$Q10:U=95.5,p=0.056,r=0.379$).
Embedded visualisation could therefore be considered easier to be picked up and faster by most people than the SP visualisation from the results of Q7 (Q7: SP\textsubscript{average}=3.67, E\textsubscript{average}=4.64).
Situated projected, according to questions 4 and 10, could also indicate that they would need more assistance or benefit from having more prior knowledge before using the system (Q4: SP\textsubscript{average}=2.58, E\textsubscript{average}=1.64;Q10: SP\textsubscript{average}=2.42, E\textsubscript{average}=1.64).
The results are presented on Figure~\ref{fig:sus_chart}.

\section{Discussion}
\label{discussion}

In this section, we analyse the obtained results in light of our hypotheses all whilst relating them to related works.

We did not find any evidence conducive of difference in efficiency from overall task completion times between E and SP formats, rejecting our H\textsubscript{1}. 
This suggests that both formats were similarly effective in terms of the time required to complete tasks, with the exception of a slight difference for tasks in the precision category. 
However, participants using E visualisations spent more time looking at the manikin, whereas those using SP visualisations spent more time consulting the projected instructions.
The format in which the information is displayed in the SP modality leads to a sharing of the participant's attentional resources for the correct execution of the task.
This behaviour illustrates that reading and executing actions involves constant interaction between the instructions and the area where the action is performed~\cite{BADDELEY197447}.
For example, participants may have needed to consult the instructions several times to ensure that the tasks were correct, which is a natural process when carrying out complex tasks.
These frequent gaze shifts observed with SP visualisations might be attributed to the attention-diverting nature of the projected virtual patient or perhaps the dynamic animations of the ECA, consistent with the concept of dynamic AR guidance~\cite{feiner1993knowledge,wang2022comprehensive}.
The lack of differences in task completion times might also stem from the pure visualisation approach without physical interactions.
Physical interactions are crucial for understanding and performing complex tasks and could have compensated for the potentially complex or hard-to-understand tasks presented solely through subtitles and oral instructions~\cite{norman1986user}. 
Future research should explore the integration of interactive elements to potentially reduce task completion times and enhance user engagement.
Exploring the potential applications of ECAs in the field similar to the meta-human represented expert applied in AR-Coach~\cite{ebnali2022ar} and similar additional tools (POCUS monitors in their case) are also relevant as reported by their multi-disciplinary subject matter expert panel~\cite{ebnali2022ar}.

In terms of precision, we observed that precision was higher for certain tasks depending on the format, partially rejecting our H\textsubscript{2}. 
Specifically, tasks involving larger target areas, such as electrode placement, were performed more precisely with E visualisations, while tasks requiring high precision in smaller areas, such as fingertip palpations, showed better results with SP visualisations. 
This finding can be attributed to potential limitations in our visual tracking system, which sometimes caused virtual elements to appear unstable and become misaligned with their physical referents. 
This flaw could cause the user to be disconnected from the scene due to conflicting visual elements, leading to them having less EWK and therefore less spatial cognition and depth perception.
This is in line with the work done by Lee~\textit{et al.}~\cite{lee2023design} mentioning the that referent size leads to issues, especially for E visualisations.
This can imply, in terms of precision, that situated visualisation on a projected virtual object may have offered a more reliable visual reference, mitigating the impact of any tracking system flaws. 
On the other hand, E visualisation could allow for more accurate object placement due to the direct alignment with physical spaces, enhancing spatial perception~\cite{azuma1997survey} but only in the case where the visualised area is larger than the potential error otherwise the indicated area would not correspond to the correct one and not be of use.
This suggests that while E visualisations provide higher precision for broader tasks, SP visualizations might be more reliable for tasks requiring fine motor control as the accuracy, in this format, is not limited by the technology.
Some of the qualitative results from the post-questionnaires interview also brought up the fact that the misalignment, was found to be a bit disturbing as they were sometimes unsure if the area was precise or not, which aligns with our thoughts.
On the other hand the SP format participants were more confused by some of the technical words mentioned such as the areas to perform the fingertip palpations.

Finally, as for the usability of our system, the results from the NASA-TLX and SUS supported our H\textsubscript{3} showing that participants experienced lower cognitive load and rated the E format as easier to use.
We observed some weak evidence that a lesser amount of external aid or prior knowledge needed for the E format.
These results can be linked to the difference in attention shifts as indicated in section~\ref{efficiency} which is the most likely cause of this higher workload.
The higher workload can also be linked back to the interviews and one of the previously brought up points being that some of the SP format participants were sometimes held up by some of the technical words used by the instructions yet were able to perform them because of the visual aid. As the participants using the E format did not bring up this fact this indicates that the SP format can make more technically complicated tasks more intuitive. 
Participants reported that the embedded instructions required less mental effort to follow and were less distracting, aligning with the cognitive load theory, which suggests that integrated and aligned visual information simplifies the cognitive process of merging multiple sources ~\cite{Hart1988DevelopmentON}.
Qualitative feedback reinforced these findings, highlighting that participants found the SP visualisations intuitive and less intrusive. 
These results suggest that better alignment and integration of AR elements with physical tasks can enhance the user experience and reduce mental strain, supporting theories that emphasize the importance of reducing cognitive effort through seamless information integration~\cite{sweller1988cognitive}.
A solution for better aligned information would be using body-aligned information for features such as the medical monitor, similar to that of the MARSOP project~\cite{markov2013impact} which showed that it reduced the amount of errors by 50\% compared to physical surface aligned information. 
Having some static non-intrusive information such as with the MobiPV wearable~\cite{boyd2016mobipv} could also be beneficial in this sense and a future development for this study.

We note that the results of our study present a few limitations such as the lack of a control batch and also a greater total batch size for increased statistic power.
These are points of interest when furthering this study.

\section{Conclusion and Future work}
\label{conclusion}

This paper contributes a comparison between using embedded information in place of situated projected information.
embedded visualizations were found to be less mentally demanding and easier to use, especially for tasks involving broader spatial alignment. Conversely, situated projected visualizations were found to be better for tasks with smaller visual markers due to their more stable visual references.
We show that this difference in user precision exists, however it is highly dependent on the quality of the displayed virtual information. Should the virtual information be misaligned for high precision tasks, we recommend situated projected information. On the other hand embedded information proved to be superior for slightly larger highlighted areas. 
A future prospective of study would be the addition of interactivity to the project which could highly benefit the user and would possibly be a better approach to reducing task completion time.
These results contribute to the growing body of knowledge on SitA and in the future, we plan to explore the integration of a more interactive system to improve the task efficiency and the application of our system in real environments, in collaboration with our national space agency to test its effectiveness in the field.

\begin{acks}
This research was first supported by the French National Center of Space Studies (CNES), VR-MARS and JUMANGI projects, Laerdal Medical and ENIB.
It was also supported by French government funding managed by the National Research Agency (ANR) under the Investments for the Future program (PIA) grant ANR-21-ESRE-0030 (CONTINUUM). 
We would like to thank the many participants who took part in our study and for providing insightful feedback which also helps us in the future development of this research, the CERV and reviewers for their feedback.
\end{acks}




\bibliographystyle{ACM-Reference-Format}
\bibliography{bibliography}

\end{document}